\documentclass[a4paper,12pt]{article}

\usepackage{ifpdf}

\newif\ifpdf
\ifx\pdfoutput\undefined
  \pdffalse
\else
  \pdfoutput=1
  \pdftrue
\fi

\RequirePackage{xspace} %
\RequirePackage{subfigure} %
\RequirePackage[centertags]{amsmath} %
\RequirePackage{amssymb}
\RequirePackage{wrapfig} %
\RequirePackage{calc} %
\RequirePackage{ifthen}
\RequirePackage{tabularx} %
\RequirePackage{flafter} %
\RequirePackage{fancyhdr} %

\ifpdf
  \RequirePackage[pdftex]{color}%
  \RequirePackage{colortbl}%
  \RequirePackage{array}%
  \RequirePackage[pdftex]{graphicx}

  \RequirePackage[ pdftex, plainpages = false, pdfpagelabels,
                 pdfpagelayout = useoutlines,
                 bookmarks,
                 breaklinks = true,
                 linktocpage,
                 pagebackref,                      
                 colorlinks = true,
                 linkcolor = blue,
                 urlcolor  = blue,
                 citecolor = blue,
                 anchorcolor = blue,
                 hyperindex = true,
                 hyperfigures
                 ]{hyperref}

\else
  \RequirePackage{color}
  \RequirePackage{colortbl}
   \RequirePackage{array}
  \RequirePackage[dvips]{graphicx}
  \RequirePackage{hyperref}
  \usepackage{rotating}
\fi

\usepackage{makeidx} 
\usepackage{setspace} 
\usepackage{rotating} 
\usepackage{ecltree}
\usepackage{epic}
\usepackage{supertabular}  
\usepackage{color}
\usepackage{exscale}
\usepackage{fontenc}
\usepackage{ifthen}
\usepackage{latexsym}
\usepackage{makeidx}
\usepackage{syntonly}
\usepackage{inputenc}
\usepackage{graphicx}
\usepackage{setspace}
\usepackage{caption2}
\usepackage[english]{babel}
\usepackage[square, comma,numbers,sort&compress]{natbib}
\usepackage{hypernat}
\usepackage{boxedminipage}
\usepackage{framed}
\usepackage{longtable}
\usepackage[all]{hypcap}

\newcommand{\note}[1]{\marginpar[left]{\singlespace \tiny #1}}

\renewcommand{\sectionmark}[1]%
      {\markright{\thesection\ #1}} 

\renewcommand{\note}[1]{}

\newcommand{\etal}     {{\it et al.}}
\newcommand{\pois}{Poiseuille}

\doublespace 

\title
{ %
\vspace*{3.0cm} \LARGE{\bf Fluid Flow at Branching Junctions} \vspace*{4.0cm} \\
}

\author{Taha Sochi\footnote{University College London, Department of Physics \& Astronomy, Gower Street, London, WC1E 6BT.
Email: t.sochi@ucl.ac.uk.} \vspace*{5.0cm}}

\begin{document}

\maketitle %
\pagenumbering{arabic}

\newpage
\phantomsection \addcontentsline{toc}{section}{Contents} %
\tableofcontents

\newpage
\phantomsection \addcontentsline{toc}{section}{Abstract} \noindent
{\noindent \LARGE \bf Abstract} \vspace{0.5cm}\\
\noindent %

The flow of fluids at branching junctions plays important kinematic and dynamic roles in most
biological and industrial flow systems. The present paper highlights some key issues related to the
flow of fluids at these junctions with special emphasis on the biological flow networks
particularly blood transportation vasculature.


\pagestyle{headings} %
\addtolength{\headheight}{+1.6pt}
\lhead[{Chapter \thechapter \thepage}]%
      {{\bfseries\rightmark}}
\rhead[{\bfseries\leftmark}]%
     {{\bfseries\thepage}} 
\headsep = 1.0cm               

\newpage
\section{Introduction} \label{Introduction}

The flow at branching junctions is a common feature of most fluid dynamics systems both in the
natural and synthetic worlds. In particular, it is one of the distinctive features of the
biological flow networks such as blood circulation and air respiration systems. There are many
studies related to the various aspects of branching flow; the majority of these studies come from
the biological and biomedical literature particularly blood flow. The obvious reason is the
vitality of this field and the importance of the role that branching flow plays in biological
systems both in normal and pathological conditions. In fact branching flow is at the foundation of
most biological systems where its importance and common occurrence can hardly be matched by any
non-biological system.

It is customary to classify the flow ducts connected to the branching junctions using the labels
`parents' and `daughters' or other synonymous words. However the definition of parent and daughter
may be rather artificial, especially for highly symmetric branching trees or chaotic flow systems,
and can be based on a geometric criterion such as the size of the vessels, where the parent is
identified as the bigger in size, or based on a flow dynamics criterion such as the flow direction
where the parent is identified as the source of flow that injects fluid into the junction. One or
both of these labeling criteria may not be applicable in some circumstances. For example, the flow
dynamics criterion for defining the parent and daughter will be the only available one for the
wholly symmetric branching where all branches are identical in size and shape with a constant angle
in between. In fact even this dynamically-based definition will not be applicable when the flow
direction is time-dependent or chaotic or ambiguous. The arbitrariness of these labels may be
highlighted by comparing the arterial part to the venous part of the blood vasculature where the
role of the parent and daughter in one of these parts according to one criterion becomes the
opposite in the other part.

For the fractal-type networks and the networks with a high degree of regularity and hence have a
strong similarity with the fractal networks, the geometric designation of parent and daughter is
reasonably clear and natural, but this becomes arbitrary in the highly irregular networks. Anyway,
there are no natural physical principles associated with these labels as such and hence the
labeling can be flexible and dependent on the context and convenience although the size criterion
seems to be more suitable in most cases. To avoid ambiguity and unnecessary phrasing complexities,
the parent and daughter in this paper will follow the commonly-used labeling which is generally
based on a branching configuration with a large parent and small daughters where the flow is in a
diverging (i.e. parent to daughters) rather than merging (i.e. daughters to parent) state.

Various types of branching occur in biological and non-biological flow networks. These types
include one-to-two branching (bifurcation) and one-to-many branching (e.g. trifurcation,
quadfurcation and so on). Branching types should also be extended technically to include more than
one parent and less than two daughters although some of these types cannot be strictly called
`branching' considering the primary meaning of the word although they have all the fluid dynamics
features of branching flow. Branching types could also be extended to include one-to-one
`branching', which occurs through an abrupt change in the shape or size (expansion or contraction)
of the flow duct, since this extension is sensible and useful in some circumstances. For example,
this extension enables flexible modeling of discontinuous transitions between two neighboring flow
ducts with different shape and/or size of their cross sectional area \cite{SochiTechnical1D2013}. A
bifurcation-type flow may occur at such one-to-one transition junctions, such as that of the
traditional benchmark problem of the 1:4 expansion, where symmetric and asymmetric branching flow
patterns occur in various flow regimes some of which are notably observed with non-Newtonian fluids
\cite{YooSR1996, ManicaB2003, RochaPO2007}.

Most branching flow studies are dedicated to the bifurcation branching in the diverging flow state.
Although the bifurcation geometry is the most common type of branching in the fluid systems,
especially in the living organisms, the branching flow in the merging state is as common as the
diverging state, e.g. the venous flow opposite to the arterial flow in the blood vasculature, and
exhalation versus inhalation in the lung airways. The geometric analogy in these two branching flow
types does not imply an analogy in the flow patterns as these two flow types can be very different.

The branching could also be symmetric or asymmetric with respect to the branching angle or radius
or both. A proposed metric for quantifying the degree of branching asymmetry in the bifurcation
case with regard to the radius size may be defined as the ratio of the radius of the small daughter
to the radius of the large daughter \cite{SchreinerNNKER1997}. This `asymmetry index' which
approaches zero for highly asymmetric bifurcation and takes the value unity for the symmetric
bifurcation is yet to be matched by a similar asymmetry measure with regard to the branching angle.
This can be simply done by replacing the radius with the angle in the previous definition, that is
the asymmetry index with regard to the bifurcation angle is the ratio of the small to the large
angles of the two daughters with respect to the extended axis of the parent. Extensions are also
required for the symmetry and asymmetry of branching with regard to the radius and angle for
non-bifurcation branching types.

In the biological flow systems, most branching is either symmetric or quasi-symmetric with regard
to the daughter radius and branching angle; or at least it does not deviate from symmetry
excessively except in exceptional cases \cite{SchreinerNNKER1997}. In fact considerable parts of
the biological flow networks, such as blood vasculature and lung wind pipes, closely match
fractal-type networks \cite{Masters2004, FinetGPRMe2008}. This may be more evident in the lung wind
pipes than in the blood vasculature due to the high symmetry of the lung compared to other organs
and porous tissue. In fact radius branching rules, as exemplified by the Murray-type laws, as given
by the forthcoming Equation \ref{MurrayEq} for a parametric branching exponent, are generally based
on the fractal nature of the branching trees. Fractal-type fluid transportation branching networks,
which are very common in nature, may be favored and naturally selected for their property of
minimizing energy consumption in comparison to other branching patterns \cite{BennettEPRNe2001,
GafiychukL2001}. The branching can also vary in shape such as T-junction, Y-junction, cross
junction, and star junction for regular branching as well as many other shapes for irregular
branching. Extensive studies about the branching flow related to most of these branching shapes can
be found in the literature (e.g. \cite{SamagaioV1989, NishimuraNH1987, MijovicL2003,
SmithPDJOe2003, WangWZGZe2008, JadeS2008, MatosO2008, LiuC2009, NematollahiNAH2009, RochaPAO2009,
KimuraOK2010, KesserwaniVRLTe2010, LiuL2011, BenesLK2012, MatosO2013, SinghSS2013}).

It is obvious that branching networks are not limited to those with cylindrically shaped tubes and
hence they include other flow duct geometries, such as ducts with square cross sectional shape
\cite{SerrenhoM2013}, and even free surface open channel networks \cite{KesserwaniVRLTe2010}. They
should also include mixed branching shapes where some branches are circular in shape while others
are triangular or square for instance. However, the analysis of those types of branching is usually
more difficult. Branching flow studies in the literature are mostly dedicated to the common type of
branching, that is confined flow networks with cylindrically shaped tubes over the whole network,
due to their common occurrence, especially in the biological flow systems, and relative ease of
modeling and analysis compared to other branching types.

Although the scope of this paper is restricted to the flow of fluids, the branching flow rules can
be extended to other types of flow, such as the electric current in the electric power networks and
electronic circuits, as there are many similarities between the two flows. In fact most of the
established branching flow laws in these two types of network have the same form and hence they are
mathematically equivalent. For example \pois\ equation is mathematically identical in form to the
Ohm's law, while the Kirchhoff's current law is identical to the continuity equation
\cite{Adam2011, PasseriniLFQV2009}. This is reflected by the exploitation of the analogy between
fluid flow and electric current in many fluid mechanical studies through the use of electric models
to describe and simulate the fluid systems as typified by the common use of the Windkessel model in
the hemodynamic investigations or the use of Kirchhoff's laws in the fluid dynamics simulations.

With regard to the blood circulation system, particularly in large mammals, which is one of the
main investigation fields of branching flow and is one of the principal subjects of the present
paper, the flow of blood in large vessels is essentially laminar with possible superposition of
minor secondary flows at branching, bending and curving zones \cite{Ku1997}. In small vessels and
capillaries the flow generally slows down and hence it steadily approaches a creeping condition in
the direction of branching of the vascular tree, i.e. in the flow direction in the arterial system
and opposite to the flow in the venous system. In fact this is a consequence of the increase in the
cross sectional area at the transition from parent to daughters. The area increase is one of the
implications of Murray-type laws with the exponent being greater than 2 \cite{Murray1926a,
Murray1926c}, as will be discussed in the forthcoming sections. However, a Murray-type law is a
sufficient condition for the area increase but not a necessary one. A consequence of this area
increase and subsequent flow slowing down is that non-Newtonian effects associated with low and
medium deformation rates become increasingly important in the direction of branching
\cite{SochiFeature2010, SochiNonNewtBlood2013}.

\section{Branch Flow Modeling} \label{Modeling}

The flow at the branching zones is normally assumed incompressible with no tangential or normal
fluid velocity at the vessel wall and with a \pois-type parabolic flow profile for Newtonian fluids
\cite{SamagaioV1989, ComerKK2001, AlharbiPC2003, LiuSZ2003, HuoK2007, SochiSlip2011, LeeCY2012}.
With regard to the blood flow, the latter condition can only be justified for the large and medium
size vessels because non-Newtonian effects become significant in the small vessels and porous
tissue resulting in a more flattened velocity profile due to the shear-thinning nature of blood
\cite{XuC1994, GijsenVJ1999, SousaCA2012, MatosO2013}.

Simplifying conditions such as fully developed flow \cite{ChenL2006} at the entrance of the outflow
vessels (daughters in diverging flow and parent in merging flow) may also be assumed although this
is not strictly valid in most cases. Less severe edge effects are also expected to occur at the
exit of inflow vessels (parent in diverging flow and daughters in merging flow). The boundary layer
at the branching zone is found to be thinner than that at the vessel wall in a fully developed flow
\cite{FresconiWP2003}.

The fully developed flow assumption could suffer further violation when complex flow patterns, such
as vortices, are induced near the junction with possible propagation by viscous diffusion; moreover
it becomes less realistic for short tubes as the edge effects become more significant in such
tubes. The assumption therefore can be justified for long tubes when the flow is laminar at
relatively low Reynolds numbers as the flow settles to its fully developed state over a short
distance \cite{BowlesDPS2005}. A Forchheimer correction term may be added to the flow equation to
account for the deviation from the parabolic profile \cite{GabrysRK2006}. With regard to the short
tubes, the fractal nature of the biological flow networks, as typified by the circulatory and
respiratory systems, observes a persistent proportionality between the length and radius of the
flow ducts making this an exception.

Other assumptions used to simplify the branching flow analysis in the biological flow networks
include the shape of the vessels as being cylindrical and the type of the flow as being laminar
uniaxial with no energy losses. Deviation from most of these assumptions are more grave at the
vessels periphery in the immediate neighborhood of the branching junctions.

In most fluid dynamics flow models, which include \pois\ and one-dimensional Navier-Stokes, the
branching junctions are assumed, explicitly or implicitly, as connecting geometric points with no
volume to store fluid or entail additional pressure loss \cite{SorbieCJ1989, Sorbie19902,
Sorbiebook1991, MabotuwanaCP2007}. This assumption simplifies the analytical and numerical
treatment of the flow in the network and splits it into a flow in tubes associated with tuning
conditions at the branching points. However, this is a gross approximation that discards many flow
features associated with the flow patterns at the branching zones which significantly contribute to
the flow in the network and define its overall behavior \cite{Lew1971}.

The \pois\ and one-dimensional models in their basic forms are indifferent to the branching angle,
and this is one of their major limitations, although angle-dependent dissipation effects have been
included in some one-dimensional models to account for this deficiency \cite{PasseriniLFQV2009}. In
fact these models lack most geometric features, such as directionality, that affect the flow in
real three-dimensional networks \cite{SochiTechnical1D2013}. This may be acceptable for the flow at
low Reynolds numbers where the process is essentially viscous, but it becomes less accurate at high
Reynolds numbers where inertial effects become increasingly important. However, in most cases of
biological flow this approximation is not far from reality and hence is generally acceptable. It
should be remarked that in this context, \pois\ flow network model should be extended to include
\pois-like non-Newtonian flow network models which are based on extending \pois\ flow network
formulation to include non-Newtonian viscous rheology for generalized Newtonian fluids
\cite{Lopezthesis2004, Balhoffthesis2005, Sochithesis2007, SochiB2008, SochiVE2009, SochiYield2010,
SochiComp2010}.

Network flow models should provide coupling conditions at the branching junctions for the flow to
be consistent. The purpose of these conditions is to force the flow in the individual ducts, which
is subject to the assumed tube flow model, to comply with the flow state in the network as a whole
according to certain coordination rules. The nature of the coupling conditions may be related to
the flow model used to describe the flow in the individual tubes that comprise the network,
although in most cases it is derived from general conservation principles such as mass and energy
conservation. The number of the required coupling conditions depends on the number of variables
used to describe the flow at the branching junctions.

More specifically, there are two prominent and widely used prototypes in the network flow
investigations, and hemodynamic studies in particular, for modeling and simulating the flow in the
fluid transportation networks: the \pois\ model for the flow in rigid tubes, and the
one-dimensional Navier-Stokes model for the flow in distensible tubes \cite{FormaggiaLQ2003,
SherwinFPP2003, HuoK2007, GeorgPH2011, SochiPois1DComp2013}. These models have different versions
with different implementations and flavors; the following description mostly applies to these
models as described in references \cite{SochiTechnical1D2013, SochiPois1DComp2013}. For the \pois\
flow network model, each junction has a single pressure variable and hence a single coupling
condition, which is normally derived from the continuity of the volumetric flow rate, as given by
Equation~\ref{ConQ}, and is based on the conservation of mass is employed. For the commonly used
form of the one-dimensional Navier-Stokes biological flow network model \cite{FormaggiaLQ2003,
SherwinFPP2003, SochiTechnical1D2013, SochiPois1DComp2013}, at each branching junction connecting
$N$ vessels there are $N$ pressure variables and $N$ flux variables and hence $2N$ coupling
constraints are required to match the flow at the branching nodes. These $2N$ constraints are
usually provided by $N$ compatibility conditions, derived from Riemann's method of characteristics
and arise from projecting the differential equations of the flow model in the direction of the
outgoing characteristics \cite{SochiTechnical1D2013}, and $N$ matching conditions based on the flow
continuity, which is derived from the conservation of mass for incompressible flow, and the
Bernoulli condition, which is based on the conservation of energy, as summarized in the following
relations
\begin{equation}\label{ConQ}
\sum_{i=1}^{n}q_{i}=0
\end{equation}
and

\begin{equation}\label{Bernoulli}
p_{k}+\frac{1}{2}\rho v_{k}^{2}-p_{l}-\frac{1}{2}\rho v_{l}^{2}=0
\end{equation}

In the last two equations, $q$ is the volumetric flow rate which is signed ($+/-$) according to its
direction (toward the junction or away from it), $i$ is a dummy index that runs over all the $n$
tubes connected to the junction, $p$ is the local pressure, $\rho$ is the fluid mass density, $v$
($=\frac{|q|}{A}$) is the fluid speed averaged over the vessel cross section, and $k$ and $l$ are
indices of two distinct branching tubes. More details about the branching coupling conditions
related to the \pois\ and one-dimensional Navier-Stokes models can be found in references
\cite{SochiTechnical1D2013, SochiPois1DComp2013}.

The use of the Bernoulli equation, which is based on the conservation of mechanical energy, is
justified by the fact that the energy losses at the junctions are normally negligible
\cite{OlufsenPKPNL2000, HuoK2007} although this assumption may not be applicable in some situations
with the involvement of complex flow patterns such as the setting of vortices and turbulence at the
junctions. The energy losses associated with the flow patterns at the branching junctions depend on
several factors such as the Reynolds number and the angle of branching \cite{HuoK2007}. A
compensation term may be added to the Bernoulli equation to account for these losses
\cite{OlufsenPKPNL2000}.

The Bernoulli condition, in the form given by Equation~\ref{Bernoulli}, is based on the assumption
of negligible gravitational body forces at the branching scale relative to the other forces
involved in the flow and hence the gravitational term in the general Bernoulli equation is dropped.
An even more reduced form of the Bernoulli equation may be used as a matching condition where only
the pressure term is maintained. The problem with the latter form is that it is valid only when the
velocity term is negligible compared to the pressure term, i.e. in the creeping flow condition,
which is not legitimate in most cases related to the biological flows where this reduced form has
been employed. A pressure continuity condition, equivalent to the latter Bernoulli condition, has
been used for the branch flow coupling in the one-dimensional flow model for networks of compliant
vessels \cite{OlufsenPKPNL2000, Olufsen2001}. However, the theoretical justification of this
condition may not be based on the Bernoulli principle in its more reduced form although they are
practically equivalent. The more reduced form of the Bernoulli condition should receive more
justification when the total area of the daughters approaches the area of the parent, which is the
case for instance in the Murray-type laws with the branching exponent approaching the
area-preserving value of 2.

Anyway, there is a general problem in using Bernoulli as a coupling condition in the
one-dimensional flow model because Bernoulli equation is based on an inviscid flow assumption which
contradicts the viscous assumption that the one-dimensional model relies upon. However, apart from
this basically conceptual contradiction the viscous effects at the junction in the network flow are
relatively small in most circumstances and hence the use of the inviscid Bernoulli condition can be
justified. The fact that the boundary layer at the branching zone is thinner, as indicated earlier,
could provide a further justification for the use of the Bernoulli equation.

The Bernoulli equation is based on other simplifying assumptions such as steady, laminar, adiabatic
flow in straight, rigid tubes \cite{OlufsenPKPNL2000}. With regard to the biological systems, and
blood distribution networks in particular where the one-dimensional model is commonly used, some of
these assumptions are not far from reality as applied in the close proximity of the branching
zones. Anyway, most of the simplifications related to the coupling conditions at the branching
zones are diluted by the more significant simplifications which are normally employed in the
principal flow models that are used to describe the flow in the vessels themselves. For example,
\pois\ equation is based on several simplifying assumptions when applied to the blood circulation,
such as laminar steady flow in straight rigid tubes. The errors associated with these
simplifications, most of which are similar to the previous coupling simplifications, are normally
more important than the errors of the coupling simplifications. Similarly the one-dimensional
Navier-Stokes distensible model disregards most of the phenomena associated with the
three-dimensional flow such as directionality, vortices, turbulence and flow separation. Moreover,
it is normally based on a pure elastic pressure-area constitutive relation instead of the more
elaborate and realistic viscoelastic characteristics although elastic models are acceptable for
describing the wall behavior in large vessels.

Another potential inconsistency in the one-dimensional flow model at the branching junctions, which
also originates from the use of the Bernoulli equation, is that according to some coupling
strategies as outlined earlier there is no unique pressure value at each junction, because the
pressure at the junctions is vessel dependent and hence there is a number of pressure values
matching the number of vessels connected to that junction. This may contradict the assumption of
zero-volume junctions and negligible viscous and non-viscous losses at the branching zones as these
assumptions are mostly based on negligible branching volume. In general, it may be argued that
while some of these assumptions are based on negligible size of the junction space, such as
negligible viscous dissipation as implied by Bernoulli, others, like non-unique pressure at the
junctions, may only be justified if the junction space is assumed sizeable.

Concerning blood perfusion in porous tissue in the microcirculation system which consists of
strongly bifurcating networks, the distensible Darcy model \cite{HuygheACR1992, KhaledV2003,
ChapelleGMC2010} is widely used to describe the flow. Because this lumped porous media model lacks
the necessary details to account for the branching effects, no branching flow features are
considered explicitly or implicitly in this model. However, very generic branching effects may
still be present within the porous medium mechanisms through intrinsic lumped parameters like
permeability; otherwise the use of these models should only be justified if minimal branching
effects are assumed in the flow regimes to which these porous models apply.

It should be remarked that reliable description of the branching flow requires three-dimensional
rather than one-dimensional models to account for the flow features that can only be accessed
through three-dimensional models. However, this may not be affordable in most circumstances due to
practical restrictions on the computational resources, especially when dealing with extensive flow
networks, as well as mathematical and numerical difficulties. The one-dimensional models then
require improvement to incorporate the essential flow features at the branching junctions to avoid
the widely adopted approach in the one-dimensional modeling where these models are centered on the
flow in the network ducts while the branching flow conditions are used only to couple the flow in
the individual tubes to orchestrate the flow in the network as a whole.

With regard to the branching in the blood vasculature, modeling and analysis of branching flow is
generally more difficult in the venous system compared to the arterial system due to the fact that
the pressure is low and hence the cross sectional area of the venous vessels could strongly deviate
from cylindrical to elliptical shape or even a collapsed state especially in the low pressure phase
of the cardiac cycle. Moreover, the presence of valves in the veins introduces more complications
on the velocity profile and flow patterns.

\section{Branching Radius}

A large number of studies are dedicated to the effect and optimal design of the radius branching
ratio especially in the biological flow systems. There seems to be a widespread consensus that
branching morphology in the biological systems is subject to optimization principles which may be
justified by evolutionary morphogenetic arguments based on natural selection \cite{Murray1926a,
Murray1926b, Murray1926c, Sherman1981, RossittiL1993a, RossittiL1993b, Rossitti1995,
EmersonCGB2006}. Radius branching rules, similar to those derived from the optimization principles,
have been observed in various biological flow systems across the animal and plant kingdoms such as
lung airways in mammals, air diffusion systems in insects, and sap transport networks in trees
\cite{RossittiL1993a, Rossitti1995}. This may originate from their close resemblance to the fractal
structures which are widespread in nature. Similar optimization arguments can also be employed to
justify the optimal design of branching networks in non-biological flow systems although simpler
physical rules can be used in the latter case. For example, biologically inspired arguments, based
on Murray's law, have been used to justify the optimal design of artificial microfluidic networks
\cite{EmersonCGB2006, BarberE2008}.

In fact there have been proposals \cite{Sherman1981} that Murray's law and the argument on which it
is based hold for any branching flow system, living or non-living, that is subject to the flow
resistance minimization objective within a specified volume. Some biological arguments cannot be
extended automatically to non-biological systems due to the involvement of biologically-specific
parameters such as metabolism although similar parameters like manufacturing cost may be applied.

The most prominent biological radius branching model is the Murray's law \cite{Murray1926a,
Murray1926c} which is based on an optimization principle related to minimizing the energy
consumption of flow systems in living organisms \cite{HutchinsMB1976}. The roots of Murray's law
can be traced back to Thomas Young and other scientists in the 19th century and the early 20th
century \cite{Young1808, Young1809, Sherman1981, ChangiziC2000, GafiychukL2001, Kizilova2008}. The
essence of Murray argument is that the radius branching morphology in the blood circulation network
is subject to an energy optimization principle where two energy consumption factors do compete: the
metabolic energy required to maintain the volume of the blood that fills the vessels, and the
mechanical energy required to pump the blood throughout the network. The energy consumption of the
first factor is directly proportional to the volume and hence to the radius squared, whereas the
consumption of the second factor is inversely proportional to the radius fourth power as a
consequence of \pois\ law \cite{Sherman1981}. While minimizing the energy consumption according to
the first factor requires diminishing the size of the blood vessels to reduce the maintained blood
volume, the second factor requires increasing the size of the vessels to reduce the flow resistance
and hence the energy of pumping. The final radius branching geometry is then determined so that the
total energy consumption required by these two factors is minimal. Murray's law has also been
explained by geometric arguments, based on the capacity of the living body for controlling blood
distribution, without resorting to the energy optimization principles \cite{GafiychukL2001}.

The pumping cost in the energy minimization argument is based on a purely viscous flow, as implied
by the use of \pois\ condition, and hence it does not account for non-viscous pumping losses. The
condition, anyway, is generally accepted in the biological networks where the flow is at relatively
low Reynolds numbers. The volume maintenance cost may also include the volume cost of the vessels
as well as the pumped fluid \cite{Sherman1981, GafiychukL2001}. Although Murray's law, and the cost
argument on which it is based, is originally derived for the blood flow where the fluid is living
and hence has a metabolic cost, it can be extended to the biological flow networks where the fluid
is inert with no metabolic cost such as the air in the respiratory system \cite{Sherman1981}. This
extension is justified by the metabolic cost associated with maintaining the living pipe network,
as indicated already, even if the extension of Murray's law to non-living flow systems is rejected.

Formally, the Murray's law is given by

\begin{equation} \label{MurrayEq}
R_{p}^{g}=\sum_{i}^{n}R_{d_{i}}^{g}
\end{equation}
where $R_{p}$ and  $R_{d_{i}}$ are the radius of the parent and the $i$th daughter vessel
respectively, $n$ is the number of daughter vessels which is 2 in most cases in the biological flow
networks, and $g$ is the branching exponent which according to Murray is 3, but other values like
2.1-2.2, 7/3, 2.6, 2.3-2.7, 2.75 and 2.0-3.0 are also theoretically derived or experimentally
observed as reported in the literature \cite{Uylings1977, Sherman1981, ChangiziC2000,
MabotuwanaCP2007, RenemanVH2009, WangLBLZe2012, LubashevskyG2013}.

\begin{figure}[!h]
\centering{}
\includegraphics
[scale=0.55] {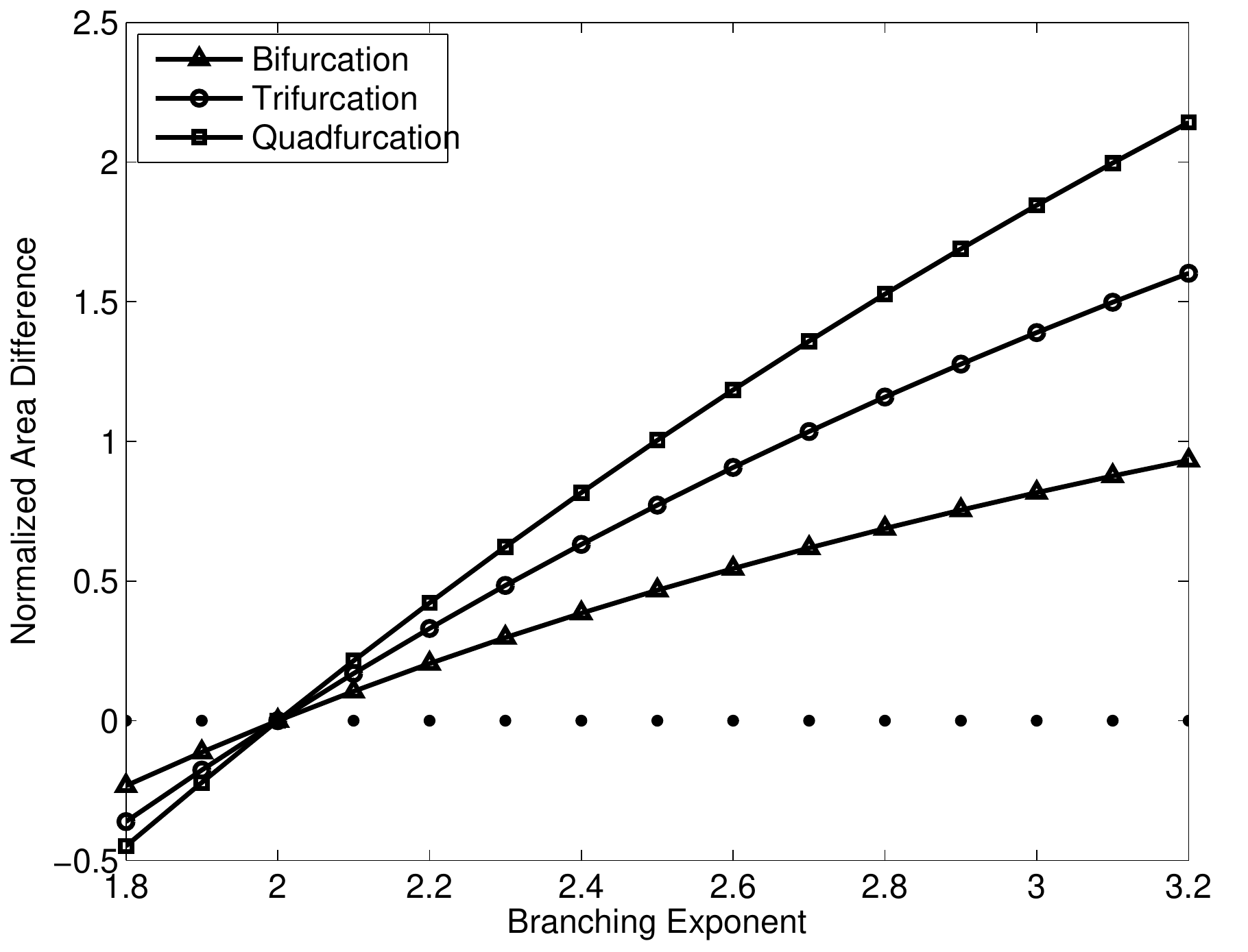} \caption{Difference between the total cross sectional area of the
daughter vessels and the cross sectional area of the parent vessel as a function of the branching
exponent for symmetric bifurcation, trifurcation and quadfurcation branching according to the
Murray-type laws where the parent and daughter radii are normalized to the parent radius.}
\label{MurrayExponent}
\end{figure}

\begin{figure}[!h]
\centering{}
\includegraphics
[scale=0.55] {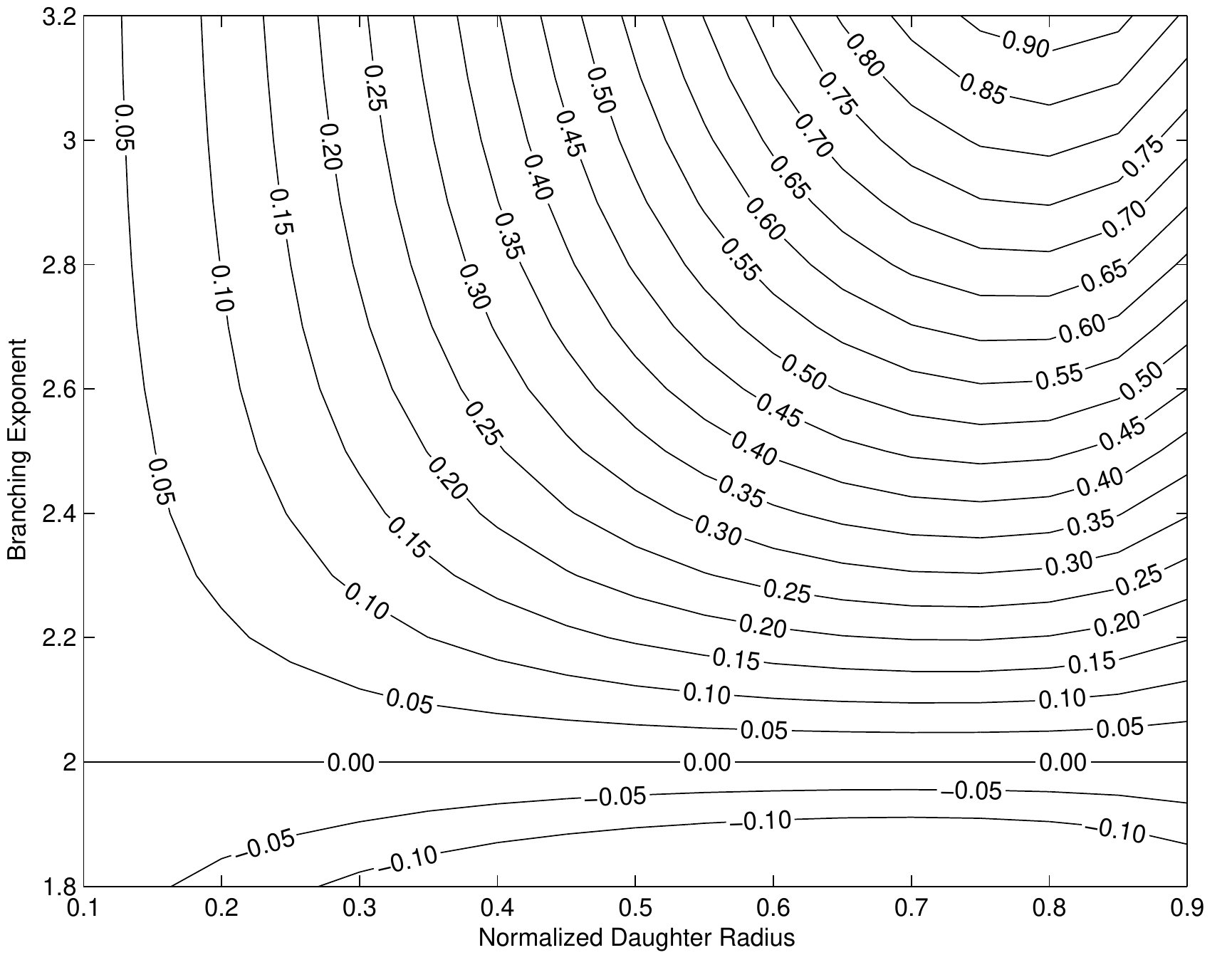} \caption{Contours of the difference between the total cross
sectional area of the daughter vessels and the cross sectional area of the parent vessel as a
function of the branching exponent and the radius of one of the daughter vessels for non-symmetric
bifurcation branching according to the Murray-type laws where all the radii are normalized to the
radius of parent. The radius of the other vessel is computed from the Murray equation
\ref{MurrayEq}.} \label{MurrayExponentCon}
\end{figure}

In Figure \ref{MurrayExponent} the difference between the total cross sectional area of the
daughter vessels and the cross sectional area of the parent vessel as a function of the branching
exponent for symmetric bifurcation, trifurcation and quadfurcation branching is plotted. As seen,
$g=2$ is the break-even area-preserving value where the parent cross sectional area is equal to the
sum of the daughters cross sectional area. Above this value, the total area of daughters exceeds
the area of parent and this difference increases as $g$ and the number of daughters increase. In
Figure \ref{MurrayExponentCon} the contours of the difference between the total cross sectional
area of the daughter vessels and the cross sectional area of the parent vessel as a function of the
branching exponent and the radius of one of the daughter vessels for non-symmetric bifurcation is
plotted. As seen in these figures, the difference in area, in favor of the daughters total area,
increases as the branching exponent increases for all the symmetric and asymmetric cases.

One of the important implications of the Murray-type laws with the branching exponent being greater
than 2 is that the total cross sectional area increases in going from one parent generation of
vessels to the next daughter generation at the branching junctions. A consequence of this increase
in the total cross sectional area is that the incompressible blood flow will slow down in the
direction from large to small vessels in the vascular network, i.e. in the flow direction in the
flow diverging networks (e.g. arterial) and opposite to this direction in the flow merging networks
(e.g. venous). This has a direct impact on several phenomena that depend on the flow speed and rate
of deformation such as pressure and non-Newtonian rheology. A consequence of this on the blood
circulation, for example, is the steady increase of the significance of shear-dependent
non-Newtonian effects, which are associated with low and medium deformation rate regimes, in the
branching direction \cite{SochiNonNewtBlood2013}.

It should be remarked that the derivation of Murray's law is based, explicitly or implicitly, on
several simplifying assumptions which include steady, laminar, Newtonian flow with a parabolic flow
profile in straight, rigid, cylindrically-shaped vessels with constant wall shear stress and
possibly identical daughters \cite{Murray1926a, Sherman1981, AlarconBM2005, MabotuwanaCP2007,
BeareDRCSe2011}. The tubes in the biological flow systems, where these models are employed, are
generally curved, tapered and distensible; moreover they are not perfectly circular in shape. The
deviation from being circular is aggravated in the blood vessels by the pulsatility of pressure
field, especially during the diastolic phase and in the venous system, where the vessels do not
only deviate significantly from being circular but can even collapse in part of the cardiac cycle.
However, most of these assumptions are widely accepted in the hemodynamic studies although some of
which, such as rigidity, are not as good approximation as others like laminar; moreover the
validity of some of these assumptions are questionable \cite{Sherman1981, ChangiziC2000,
RenemanVH2009}.

There are attempts to incorporate complex blood rheology, which includes radius dependent viscosity
and hematocrit level effects, in the Murray's formulation \cite{AlarconBM2005}. Extensions of
Murray's law to flow networks with non-cylindrical ducts have been proposed for the design of
synthetic flow systems \cite{EmersonCGB2006, BarberE2008}. Murray's law has also been extended to
include steady-state turbulent flow with the branching exponent taking the value 7/3
\cite{Uylings1977, Sherman1981}. Other efforts in this context include the generalization of
Murray's law to include non-Newtonian fluids of power law type \cite{RevellinRBB2009, Tesch2010}
and the extension to non-circular tubes with elliptical cross section \cite{Tesch2010}.

A possible problem with the Murray-type laws, which is related to the assumption of rigidity of the
branching vessels, is that due to the pulsatility of the pressure field and the flexibility of the
biological vessels, the size of the vessels usually varies during the flow cycle (e.g. cardiac or
respiratory cycle). Because this variation in general could be out of proportion to the reference
pressure magnitude by scaling the branching vessels up or down by the same factor, the radius ratio
could change throughout the cycle rendering the optimization principle, or any other principle on
which the derivation is based, invalid. However, due to the close geometric and material
similarities between the parent and daughter vessels the scaling should not be far from
proportionality throughout the whole flow cycle. The effects of propagating pressure waves that
distorts the radius ratio in selective areas and hence affecting the validity of Murray-type laws
should be minimal.

The robustness of Murray-type laws may also be undermined in the small capillaries due to the
sudden change in the blood viscosity at the junction transition caused by the
F{\aa}hr{\ae}us-Lindqvist effect and plasma skimming which results in a lower effective viscosity
in the smaller daughter vessels although this seems to have minor effect on the validity of the
Murray-type laws in general \cite{Sherman1981}. Anyway, this only applies to a limited generation
of branching vessels where F{\aa}hr{\ae}us-Lindqvist effect takes place.

As indicated earlier, some of the theoretical studies, which are based on different optimization or
geometric or purely fluid mechanical arguments, have concluded radius branching laws similar in
form to the Murray's law but with different values for the branching exponent, $g$. Area
preservation principle with $g=2$ has also been proposed as a radius branching law
\cite{BennettEZZMe2000, ChangiziC2000, GafiychukL2001, LubashevskyG2013}. Strange values ranging
between 1.2-1.6 for the branching exponent, which undermine the principle of minimum work argument
as well as some other arguments, has also been reported in relation to carotid bifurcation
\cite{BeareDRCSe2011}. Despite the fact that an exponent value of less than 2 in the Murray-type
models is possible in principle for certain types of branching in the blood network, it cannot be
accepted in general over major parts of the vascular tree due to the existing evidence in support
of the fact that the flow in general slows down as the size of the vessels converges toward the
capillary networks and porous tissue.

Other studies have led to radius branching models different in form to the Murray's law. One of
these is the empirical model proposed recently by Finet \etal\ \cite{FinetGPRMe2008} which, for a
bifurcation, is given by

\begin{equation}
R_{p}=0.678\left( R_{d_{1}} + R_{d_{2}} \right)
\end{equation}

Assuming the validity of the Murray-type laws in their general form with a parametric branching
exponent, several reasons can be proposed to explain the discrepancy in the reported values of the
exponent as observed and measured in the experimental studies. These reasons include measurement
and analysis errors associated with some vagueness in the definition of the radius of parent and
daughters, especially in the branching neighborhood, as well as tapering which is a general feature
in the biological flow networks. However, some of these discrepancies are too large to explain by
random or systematic errors. The exponent could also vary in pathological cases due to
cardiovascular diseases \cite{HutchinsMB1976} which adjust the vessels cross sectional area with or
without the deposition of foreign materials on the luminal surface. Some of these discrepancies may
also be related to the preparation and measurement techniques, as well as differences in the
applied procedural and analysis methods. Variations between species and individual subjects is
another possible reason for some of these contradictions. The branching exponent may also vary
depending on the location and rank in the vascular tree and the difference between the type of
vasculature such as arterial versus venous \cite{PriesS2005}.

Some studies suggested that Murray's law is good for large arteries and arterioles but not for
microcirculation networks \cite{AlarconBM2005}, while others seem to suggest an increasing exponent
down the arterial tree \cite{RenemanVH2009, BeareDRCSe2011}. The convergence of the branching
exponent to the theoretical Murray value of 3 with decreasing vessels size has also been reported
in one study with the explanation that \pois\ law, which most Murray-type models are based upon, is
better approximated in the small vessels \cite{LeeL2010}.

In summary, the reported results in the literature are not only infested by significant
discrepancies in the value of the branching exponent between different studies, but also by large
fluctuations and error margins in some of these studies, and this subjects the proposed Murray-type
laws to many uncertainties and question marks. Similarly, the other empirical and theoretical
radius branching models, which are different in form to the Murray-type models, are not far away
from controversies and uncertainties.

\section{Branching Angle}

The geometry of the branching zones plays an important role in the distribution and collection of
fluids and their ingredients. In the blood circulation network, the shape of branching, which
includes branching angle, has an obvious impact on the flow of blood and the movement of its
constituents, such as red blood cells, in the vessels. The branching angle of the daughter tube is
normally measured between the daughter axis and the extended orientation of the parent axis. Some
ambiguity in the definition of branching angle especially in the biological systems may arise due,
for example, to irregular shape of the parent and daughter vessels and the deviation from the
optimal cylindrical shape as well as curvature of the vessels at the branching zone. Pulsatility of
biological flow, as seen in circulation and respiration, associated with possible change in the
branch orientation and apex position \cite{RossittiL1993b} can lead to a time-dependent alteration
in the branching angle.

The definition of the branching angle in the literature is normally based on a two-dimensional
branching configuration where the axes of the branching vessels are coplanar; moreover, these
definitions are generally based on a bifurcation branching type and hence some ambiguity may arise
with branching in three-dimensional networks and branching orders higher than bifurcation. However,
the previous definition of the branching angle should apply to branching vessels that do not share
a common plane. Similarly, non-bifurcation branching can be accommodated with some flexibility in
the definition of branching angle.

Work minimization arguments, similar to the ones proposed for the radius branching ratio, have been
proposed for the optimal design of branching angle \cite{Murray1926c, RossittiL1993a}. However, it
has been suggested that branching angles are irrelevant to the minimization of energy consumption,
due to their large variations, and hence they are generally determined by other factors
\cite{HutchinsMB1976, RossittiL1993b, Rossitti1995, TadjfarS2004}. In fact a potential optimization
principle in the branching angle design does not necessarily require a fixed branching angle,
similar to the fixed radius ratio of Murray's law for instance, since optimization of any physical
parameter like energy in the flow system may require variable angle depending on other factors.
Anyway, the most important factor that determines the branching angle in the living tissue should
be the optimal delivery of the fluid to the target tissue and this seems to be the most appropriate
design principle for the branching angle.

Great variations in the branching angle are also observed in non-living natural and synthetic flow
networks, such as geological structures and manufactured porous media which are subject to
spontaneous sedimentation and flow formation processes. Although this may not be significant due to
the absence of obvious optimization principle in the formation of these systems, these processes
could be subject to certain optimization rules that include the branching angle of the flow ducts.
In fact optimization principles are at the heart of many physical phenomena in the natural and
synthetic worlds.

Branching junctions are recognized to lower the critical threshold of Reynolds number for setting
turbulence. The angle of branching has an influence on the transition from laminar to turbulent
flow where the threshold limit decreases with increasing angle \cite{RossittiL1993b}. This is due
to inertial effects where the increase in the branching angle entails larger and more abrupt change
in momentum that promotes the setting of turbulence. The bifurcation angle also has a strong
influence on the secondary flows and recirculation zones at the branching regions
\cite{AlharbiPC2003, ComerfordPD2008}. On the other hand, branching angle seems to have a minor
influence on plasma skimming, and phase separation in general. One possible reason is that these
effects occur in the minute capillaries where creeping flow of viscous nature is the norm.

Concerning the relation between the radius and angle of branching, there is no correlation between
the branching angle and the size, i.e. calibre, of the branching tree in the biological flow
networks \cite{HutchinsMB1976}. This is in concordance with the approximate fractal nature of these
networks. However, there is a tentative relation between the parent-to-daughter size ratio and the
branching angle which increases as the ratio increases; that is the smaller daughter vessel
branches at a larger angle than the larger one \cite{Adam2011}.

Despite the fact that both radius branching ratio and branching angle are influential in
determining the flow patterns at the branching regions, one of these factors may be more
influential than the other for certain phenomena or in certain circumstances. There are no general
rules about this due to the complexity of these issues and the involvement of many factors; such as
the type of branching, Reynolds regime, and fluid rheology; although some definite conclusions have
been reported in the literature in relation to some of these issues \cite{Tadjfar2006}.

With regard to the experimental methods used to investigate the radius and angle branching laws,
various imaging and measurement techniques, such as computed tomography scan, magnetic resonance
imaging and confocal laser microscopy, as well as other acquisition techniques like polymer casting
and cryo-microtome sectioning have been used to obtain geometric and topological information from
biological branching networks \cite{SchreinerNNKER1997, AharinejadSN1998, TaylorHZ1998,
ZhaoXHTSe2000, SteinmanTLMRe2002, ThomasMRS2003, YounisMCIHe2004, HunterLCZGe2006, SinnottCP2006,
CassotLLPD2009, BlancoPUF2009, GrinbergACMK2009, GeorgPH2011, CasciaroCGGA2011, WangLBLZe2012,
CardenesDDPF2013, TuPLBRW2013}. However, the majority of these techniques and their complementary
analysis methods are susceptible to significant errors and hence cannot provide accurate
geometrical data to test the validity of the branching rules such as Murray's law. For instance,
there are technical difficulties in making precise measurements of the radius around the complexly
shaped branching region due partly to technical difficulties and partly to some vagueness in its
definition at the branching zone. Similarly, there is an ambiguity in the definition of the vessels
axes, due to curvatures and shape irregularities, that prevents precise determination of the
branching angle. Moreover, most branching measurements are carried out assuming the branching trees
are contained in a two-dimensional plane, whereas in reality these trees are three-dimensional
entities.

\section{Branching Effects}

There are various flow effects that occur at the branching junctions; some of which are briefly
discussed in this section. In general, branching flow effects complicate the flow patterns and
hence flow modeling and analysis at the branching regions. These effects are either triggered or
exacerbated by the branching flow. One of these effects is the non-Newtonian rheology
\cite{SochiArticle2010, SochiNonNewtBlood2013} which can have a major contribution to the flow
patterns at the branching junctions. Non-Newtonian effects are supported by experimental and
numerical investigations \cite{NakamuraZNIH1985, BanerjeeCB1993, LouY1993, YooSR1996, GijsenVJ1999,
MijovicL2003, ChenL2004, ChenL2006, BroboanaMB2007, RochaPO2007, BarthBC2008, MatosO2008,
MatosO2013} where significant differences between the behavior of Newtonian and non-Newtonian
fluids at the branching junctions have been widely observed. Although non-Newtonian effects are
originally related to the rheology of the fluid and hence are not specific or limited to the flow
at the branching regions, the complex geometric factors at the branching zone can stimulate or
aggravate these effects. For example the converging-diverging and tortuous nature of the flow paths
in these zones can activate ceratin viscoelastic modes associated with the fluid rheology.

Another type of branching flow effects is time-dependency. There are two main reasons for the time
dependent effects to occur in the branching flow: the non-Newtonian rheology
\cite{SochiNonNewtBlood2013} such as thixotropy and viscoelasticity, and the pulsatile nature of
the flow associated with the pulsatility of the vessels size, pressure field and velocity profiles
\cite{GhalichiThesis1998} as seen for example in the blood circulation. Although these causes are
not specific to the flow at the branching junctions, some of which may be stimulated or exacerbated
by the flow types in the branching zones. The time dependent features of the branching flow can be
very complex and include recirculation zones, skewed velocity profiles, flow separation regions and
secondary flows. Some of these features also exist in the time independent non-pulsatile branching
flow \cite{MotomiyaK1984, NakamuraZNIH1985, CebralYLSMe2001, RochaPO2007, MatosO2013}. These
features and the contributing factors complicate the flow patterns at the branching junctions and
make the analysis more difficult.

Another effect, which is a distinctive feature of the branching flow, is phase separation where a
complex multi-phase fluid disintegrates into its components at the diverging junction due to the
fact that the daughter pathway is more favorable to the passage of a particular fluid phase than to
the other phases. In blood circulation, phase separation demonstrates itself in plasma skimming in
the minute capillaries with a considerable drop in the fluid viscosity due to a low level of
hematocrit in the daughter vessels which results from the reduction in the vessels size as it
becomes comparable to the size of the red blood cells.

Branching flow effects also include the occurrence of several complex flow patterns such as the
formation of turbulence zones and vortices at the branching regions with possible energy losses.
The presence of constrictions and wall deformability, due for example to a stenosis or aneurysm, at
or near the branching junction can cause the development or exacerbation of these complex flow
patterns. However in biological systems, such as blood flow in arteries and air flow in the lung
air pipes, such complex flow patterns are the exception rather than the norm due to the laminar
nature of the flow over predominant parts of the flow network. Moreover when these exceptional flow
patterns occur they normally disappear quickly due to the time-dependent pulsatile nature of the
flow although in some circumstances they could be persistent and follow the periodic pattern of the
primary flow. Other factors, such as loops, in the branching flow network can exacerbate the
situation and introduce more complications on the flow patterns at the branching regions.

\section{Circulatory Branching Lesions}\label{Lesions}

In this section, we outline some of the branching lesions which are normally found in the blood
circulation system. Despite the fact that the causes of the arterial diseases are common to all
parts of the flow system, whether branching zones or not, the branching zones are generally favored
for the development of arterial lesions, such as plaque depositions and aneurysms. This is due to
the localized complex flow and shear stress patterns and the involvement of composite hemorheologic
and hemodynamic factors; such as pressure, fluid velocity, and particle residence time; which can
modify the geometric and material properties of the vessels wall at the neighborhood of these sites
through sedimentation of foreign materials, alteration of the physical characteristics of the
original wall material, and changing the shape of the walls. In this regard, the endothelial layer,
which is highly sensitive to the magnitude and fluctuations of the wall shear stress, plays a
significant role in the genesis and progression of lesions at the branching zones and in blood
vessels in general.

The pulsatility of flow, which results in a cyclic change in the geometry of branching zones such
as the apex position and branching angle as well as the radius size, can also influence the
formation and exacerbation of branching lesions through the imposition of persistent abnormal flow
and stress patterns and fatigue zones by creating, for example, turbulence regions, or inflicting
high or low shear stress, or stretching or shrinking certain spots around the branching region.
These problems can be worsened by the involvement of other supporting factors like arterial
hypertension and aging. The rheology of blood, which essentially behaves as a Newtonian fluid in
large vessels and non-Newtonian in small vessels, can also have a positive or negative impact on
the branching lesions. For example, non-Newtonian effects can contribute to the complication of the
flow and pressure patterns at the branching zones affecting, directly or indirectly, positively or
negatively, the genesis and progression of these lesions \cite{BrienEF1976, RossittiL1993a,
RossittiL1993b, Rossitti1995, GreilPWWLe2003, ValenciaGFA2006, ValenciaLRBG2008, RauHVVDe2008,
ComerfordPD2008, GrinbergACMK2009, VosseS2011}.

Several geometric and hemodynamic factors contribute to the development and progression of lesions
at the branching zones. The contribution of most of these factors is based on effects related to
fluid-structure interaction. The flow pattern at the branching junctions, especially if associated
with complex flow phenomena such as vortices, may create time- and space-dependent wall shear
stress that contributes in the long term to the development of lesions such as the deposition of
plaque, wall thickening and reduced distensibility \cite{CollinsM1979, BanerjeeCB1993,
ArslanTND2005}. Plaques seem to develop mostly in the low endothelial wall shear stress areas at
the branching junctions and hence these areas are more susceptible to plaque-related lesions like
atherosclerosis which commonly occurs near the branching zones such as lateral bifurcation walls
\cite{MotomiyaK1984, KuGZG1985, KramsWOVSe1997, StoneCKCSe2003, StoneCYKPe2003, Arslan2006,
LouvardMHL2010}. Low endothelial shear stress may also play a role in the formation of thrombosis
and re-stenosis following stenting operation of atherosclerotic lesions \cite{KoskinasCAG2012}.

\section{Wave Propagation at Junctions}

In general, there are four main factors that affect the transmission and reflection of flow and
pressure waves in the fluid-filled tubes: the physical properties of the fluid such as mass
density, the physical properties of the tubes material such as Young's modulus, the geometry of the
tubes such as tube radius, and the fluid-structure interaction as represented for instance by the
interactive relation between the pressure and cross sectional area. The significance and
contribution of these factors vary in different circumstances; moreover the contribution could
depend on other aspects as well.

These factors are common to the branching junctions and other zones in the fluid transportation
networks. What is specific and particularly relevant to the branching zones, especially in the
context of distensible biological flow networks such as blood vasculature, is the geometry of the
junction zone and the physical characteristics of the wall material, such as the elasticity or
viscoelasticity of the wall, since these characteristics are subject to alterations due, for
example, to plaque depositions and atherogenic processes \cite{Li1986}. As discussed in section
\ref{Lesions}, the branching zones are highly susceptible to such adaptations and
characteristics-changing developments. As indicated earlier, some of the geometric and material
factors that affect wave transmission and reflection are partly reflected in the employed
constitutive relation that correlates the transmural pressure to the tube cross sectional area for
modeling the branching flow; which in essence is a fluid-structure interaction influence.

Partial reflection of the pressure wave takes place at the points of abrupt change in the vessel
shape and its total cross sectional area; the most obvious example of these reflection points are
the branching junctions. In the blood transportation vasculature, stents, stenoses and aneurysms
can also be points for wave reflection; moreover they normally affect the material properties, like
elasticity or viscoelasticity, and geometry, such as stenotic constriction or aneurysmal dilation,
of the walls resulting in alteration of the wave characteristics like the speed of propagation
\cite{VosseS2011}.

\section{Conclusions} \label{Conclusions}

Branching flow is commonplace in fluid dynamics systems in general and biological flow networks in
particular. The most prominent example of the biological models that govern radius branching is the
Murray's law and its variations. However, there are many controversies in the literature about the
validity of Murray's law as well as inconsistencies related to the value of the branching exponent
assuming the validity of the law in its generic form. While some studies categorically support the
Murray's law, others agree with this law in form only with a different value for the branching
exponent. Yet other investigations totally disagree with Murray's law since no exponent value
within acceptable error margin was found to satisfy the general form of Murray's law according to
these studies.

Assuming the physical reality of the design rules that govern radius branching ratio in the
biological and naturally-occurring systems, the problem with the proposed theoretical models, as
exemplified by Murray's law, is that there is no unique argument that can be instated to derive and
substantiate these relations due to the existence of various possibilities for the principle that
can be used to justify such relations. Therefore, these relations can only be accepted if there is
an overwhelming evidence from experimental and observational data over major parts of the branching
flow network in support of these models. As there are many inconsistencies and controversies in the
literature about the parametric values in these relations as well as their general form, these
models should be treated with caution. Regarding the empirical models, more thorough investigations
are required to establish these models. Possible variations in the form and parametric values in
the proposed radius branching models between species, individuals, and even position and rank in
the flow networks should also be considered.

Despite all these controversial issues, there is one thing that seems to be established about the
radius branching in predominant parts of the blood circulatory system, that is the total area
increases at the branching junctions in the transition from large to small vessels with an obvious
consequence that the flow generally slows down in the branching direction. However, this does not
imply a Murray-type law or any other theoretical or empirical model due to the existence of various
alternatives with regard to the form and parametric values as well as the possibility of the
absence of a persistent radius branching pattern over the whole or even major parts of these
networks. In the absence of thorough studies that cover all the biological flow networks in
different species considering possible variations in the individual subjects and their health
status, no general law can be definitely concluded.

Concerning the branching lesions in the circulation system, complex flow patterns at the branching
zones; as well as material, geometric and hemodynamic factors such as pressure, fluid velocity, and
wall shear stress; play a significant role in the genesis and progression of these lesions. The
branching zone is normally a low shear stress area, and since low endothelial shear stress is a
stimulus for plaque formation and atherogenesis the branching zones are strong candidates for
lesions like atherosclerosis and stenoses. Clinical intervention such as bifurcation stenting can
also introduce changes on the wall shear stress patterns at the branching zones with long term
consequences on the development and progression of branching lesions. Periodical change in the
shape of the branching region, such as the angle of branching and apex position, due to the flow
pulsatility can also have an impact on the development of lesions originating, for instance, from
fatigue and aneurysmal dilation. Recirculation zones in the branching region may also act as a
stimulus for the development of lesions like atherosclerosis and thrombosis. In summary, the
branching zones are exceptionally susceptible to the genesis and progression of arterial defects
due largely to the complex fluid and solid dynamics involved in these zones as well as
fluid-structure interaction factors.

Non-Newtonian rheology, whose importance increases in the small blood vessels, introduces more
complex branching flow patterns such as setting or increasing the magnitude or widening the
separation areas of vortices and turbulence zones. Non-Newtonian rheology and its ensuing effects
can also influence the localization, distribution and magnitude of the wall shear stress around the
branching junctions with possible long term impact on facilitating or hindering lesions such as
stroke. Although energy losses at the branching junctions are generally negligible in the
biological flow systems, the development of complex flow patterns, such as turbulent fluctuations
and transient and steady-state vortices, can increase their significance.

Wave reflection generally occurs at points of sudden change in the geometry of the transmission
route. Branching points therefore have a major contribution to the transmission and reflection of
the pressure waves in the distensible fluid-filled networks with obvious impact on the fluid
transportation and long term evolution of these flow systems.

\newpage
\phantomsection \addcontentsline{toc}{section}{References} %
\bibliographystyle{unsrt}

\end{document}